\documentstyle[12pt,aasms4,epsfig,rotate]{article}
\newcommand{\etal}{ {\it et al.}}

\newcommand{\msol}{M$_{\sun}$}
\newcommand{\mdot}{$\dot{M}$}

\begin{document}

\title{THE 1996 SOFT STATE TRANSITIONS OF CYGNUS~X-1}
\author{S. N. Zhang$^{1,2}$, W. Cui$^{3}$, 
B.~A.~Harmon$^{1}$, W.~S.~Paciesas$^{1,4}$,
R.E.~Remillard$^{3}$ and J.~van~Paradijs$^{4,5}$}
\affil{$^{1}$ES-84, NASA/Marshall Space Flight Center, Huntsville, AL 35812, USA\\
$^{2}$Universities Space Research Association\\
$^{3}$Center for Space Research, Massachusetts Institute of 
Technology\\
$^{4}$University of Alabama in Huntsville\\
$^{5}$University of Amsterdam and Center for High-Energy Astrophysics\\
}

\vspace{3cm}
\begin{center}
Accepted for ApJL publication on 1/4/97.
\end{center}

\abstract{
We report continuous monitoring of Cygnus X-1 in the 1.3 -- 200 keV 
band using ASM/RXTE and BATSE/CGRO for about 200 days from 1996 February 21
to 1996 early September. During this period Cygnus X-1 experienced a 
hard-to-soft and then a soft-to-hard state transition. The low-energy 
X-ray (1.3-12 keV) and high-energy X-ray (20-200 keV) 
fluxes are strongly anti-correlated during this period.
During the state transitions flux variations of about a 
factor of 5 and 15 were seen in
the 1.3-3.0 keV and
100-200 keV bands, respectively, while the average 
4.8-12 keV flux remains almost unchanged. 
The net effect of this pivoting is that 
the total 1.3-200 keV luminosity remained unchanged to within $\sim$15\%.
The bolometric luminosity in the soft state 
may be as high as 50-70\% above the hard state luminosity,
after color corrections for the luminosity
below 1.3 keV.
The blackbody component flux and temperature
increase in the soft state is probably caused by a combination of the
optically thick disk mass accretion rate increase and
a decrease of the inner disk radius.} 

 
\section{Introduction}

Cygnus~X-1 is  one of the brightest high-energy sources in the sky, with an  
average 1-200 keV energy flux of $\sim$3$\times$10$^{-8}$ erg/cm$^{2}$/sec.
Most of the time its X-ray spectrum is very hard, with 
a high-energy ($> 20$ keV) X-ray flux 
around or above that of the Crab Nebula while its low-energy ($<10$ keV) 
X-ray flux
is around 0.5 Crab. Occasionally, the spectrum of Cyg~X-1 becomes much 
softer; then its low-energy X-ray flux increases to 1 -- 2 Crab
while the hard X-ray flux decreases to 0.5 Crab or less. 
On the basis of the radial-velocity curve of its O9.7 Iab companion star, 
Webster and Murdin (1971) and Bolton (1971) concluded that the compact star in 
Cyg~X-1 may be a black hole (BH); their results have been confirmed by later 
detailed work of Gies and Bolton 
(1986) who concluded that the mass of the compact object in Cyg~X-1 
is greater than 7 \msol, but more probably
16 \msol, far exceeding the theoretical (and observational) upper mass limit 
of 3.2 \msol\
for a neutron star. It is thus the first stellar mass BH candidate
(BHC) (cf. Liang and Nolan 1984 and Lewin and Tanaka 1995, for reviews and 
references therein).

Cyg~X-1 is often considered as the canonical BHC; 
many of its characteristics, such as its high X-ray luminosity
above 100 keV, the ultra-soft component in its X-ray spectrum, the 
hard/soft X-ray flux anti-correlation, and the rapid X-ray flux variability, 
are shared by other systems, which on the 
basis of a dynamical mass estimate may contain a BH. 
It is, however, currently
not understood why accreting BHs show these characteristics. 

In spite of extensive studies over the last three decades, 
the mass accretion conditions near the central compact object in Cyg~X-1 
and other BH systems, and the high-energy 
radiation mechanism are still poorly understood. This 
remains an outstanding issue in high-energy 
astrophysics. State transitions fully covered over a large range in X-ray 
photon energy may provide us with
important clues towards a better understanding of these BH X-ray 
binaries. During a
transition, a rich collection of information may be obtained, such as flux
variations at many time scales in all energy bands, energy spectral evolution,
correlations between different energy bands, etc; such data should provide
useful tests of various theories and models. 

Previously, several transitions between the hard state (HS) and soft state (SS) of Cyg~X-1
have been
observed. In 1971, a soft-to hard (S-to-H) state transition was observed in the
2-20 keV band (Tananbaum \etal\ 1972). In 1975, a complete
transition was observed with
{\it Ariel} V between 3 and 6 keV (Holt \etal\ 1976), and by
{\it Vela} between 3 and 12 keV
(Priedhorsky \etal\ 1983). These observations were limited to rather low
X-ray energies. The only broad-band observations of a transition
were obtained in 1980, with {\it Hakucho} between 1 and 12
keV (Ogawara \etal\ 1981) and with HEAO-3
between 48 and 183 keV (Ling \etal\ 1983). However, the SS onset was
observed only with HEAO-3 as a rapid decrease in the 48-183 keV flux,
and then later the SS was observed with {\it Hakucho}
only when the 1-12 keV flux was already a
factor of 2-3 higher than its HS level before the transition.
Therefore, no simultaneous low- and high-energy
X-ray observations during S-to-H
or hard-to-soft (H-to-S) transitions of Cyg~X-1 have been made so far.

In this paper we report results obtained from the near continuous monitoring
of Cyg~X-1 during a complete state transition from the HS to the 
SS (Cui 1996; Cui, Focke and Swank 1996; Zhang \etal\ 1996a),  
and then back to the HS (Zhang \etal\ 1996b, 1996c), 
observed simultaneously with 
ASM/RXTE (1.3-12 keV) and BATSE/CGRO (20-200 keV) for about 200 days from
February to September 1996. Our results are 
complemented by detailed pointing observations
with ASCA (0.5-10 keV) (Dotani \etal\ 1996), and with the PCA and HEXTE 
(2-250 keV) aboard RXTE
(Cui, Focke zhang Swank 1996; Belloni \etal\ 1996; Cui \etal\ 1997), and with
SAX (0.1-300 keV) (Piro \etal\ 1996)
and with OSSE/CGRO
(50-600 keV) (Phlips \etal\ 1997). We will focus on variations in the total luminosity during this
period and on the correlations between the low-energy and high-energy 
X-ray fluxes. 

\section{Observational Data}

The Rossi X-ray Timing Explorer (RXTE) was launched on 1995 December 30. 
The All-Sky Monitor (ASM) began normal operation
on 1996 February 21. More than 70
sources have been routinely monitored in 
three energy bands (1.3-3.0, 3.0-4.8 and 4.8-12 keV).  A detailed
description of the performance of the ASM, as well as its
calibration and data reduction procedures, has been given by
Levine \etal\      (1996).

The BATSE experiment is one of the four instruments aboard
the Compton Gamma Ray Observatory, and has operated continuously since launch
in April 1991. BATSE can
monitor the entire hard X-ray (HXR) sky with almost uniform
sensitivity for the detection of gamma-ray bursts, solar
flares, pulsars and other persistent and transient HXR
sources (Fishman \etal\     1989). The persistent and 
transient HXR source monitoring is achieved by using
the Earth occultation technique (Harmon \etal\     1992) and Earth
occultation transform imaging technique (Zhang \etal\     1993). 
Cyg~X-1 has been one of the brightest sources in the BATSE database, in 
which it is detected from 20 to above 300 keV (see Crary \etal\ 1996, and 
Paciesas \etal\ 1996, for recent papers describing results of analyses of 
the BATSE data on Cyg~X-1).

\section{Light curves and total luminosity variations}

The daily averaged ASM light curves in the three energy bands are
plotted in figure 1 (upper panel), in units of the total Crab Nebula
counting rates detected by ASM in the same energy bins. (The three-channel
data on each day would overlap each other for an energy spectrum
with the same shape as that of the Crab Nebula (power law with a photon 
index of $\sim$ --2). For a steeper spectrum the
lower energy channel data will lie
above the higher energy data and vise versa.)
In the middle panel of the figure we plot the three energy
band (20-50, 50-100 and 100-200 keV) BATSE light curves. There are a total of
8-9 energy channels covering the 20-200 keV energy range. For each energy band, we
use the detector response matrices and integrate over the relevant energy 
channels to produce the photon fluxes, by assuming a power law shape of a
photon index of --2.0. The fluxes are then divided by the fluxes of the Crab
Nebula detected with BATSE in the same energy range. By converting the fluxes 
to `Crab' units, we avoid possible absolute detector calibration problems
between ASM and BATSE.

Two spectral state transitions occurred near TJD 10220 and
TJD 10307, characterised by the rapid increase and 
decrease of the ASM low-energy flux. We call the state ``hard state" 
before
TJD 10220 and after TJD 10307, since the overall spectral shape is harder
than that of the Crab Nebula. Similarly we call the state between the
transitions the ``soft state". The HS spectrum is characterised by
a hard power law between 1.3-100 keV with a photon index of around --1.8. 
The 20-50 keV flux is above the power law by 20-30\%. Above 100
keV, a spectral cutoff is observed. In the SS, the overall 3-200
keV spectrum is dominated by a power law with a photon index
of about --2.5. The first ASM energy band is contaminated significantly
by a soft excess. 
A flux excess between 20-50 keV above the power law, 
similar to the HS spectrum, 
makes the overall 20-200 keV spectrum fit reasonably well
with a cutoff power law model. The details of the spectral evolution will be
presented elsewhere (Zhang \etal\ 1996d)

In the bottom panel of the figure we plot the total luminosity observed with ASM
and BATSE between 1.3-200 keV. For the ASM bands, the luminosity is calculated
in each band separately by assuming the Crab Nebula spectrum shape. A 
hydrogen column density $N_{\rm H}$$\sim$5.5$\times$10$^{21}$ atoms/cm$^{2}$ (Ebisawa
\etal\ 1996) is used
for correcting the absorption in the 1.3-3.0 band. This correction is not
sensitive to the exact value of $N_{\rm H}$ we used. The luminosity would be
underestimated in the 1.3-3.0 keV only by $\sim$5\% if the true 
$N_{\rm H}$ is 7.0$\times$10$^{21}$ atoms/cm$^{2}$.
The luminosity in the BATSE energy range (20-200 keV) is
calculated by assuming that the energy spectrum is a power law with an
exponential cutoff (e.g., Sunyaev and Truemper 1979; Sunyaev and Titarchuk 1980)
(the power law index and the cutoff energy on each day were separately 
determined from spectral fits to the daily averaged count rates in 8 or 9 
spectral energy bands).
The gap between the ASM and the BATSE
energy bands is filled by interpolating the 4.8-12 keV and the 20-30 keV
fluxes and assuming again a power law energy spectrum. Overall, we estimate
that the deviations of the 
calculated luminosity should be
less than 10\% of the true luminosity in the 1.3-200 keV energy band. A
distance of 2.5 kpc is assumed for Cyg~X-1 in these calculations.

There is a general anti-correlation between the ASM fluxes and the BATSE fluxes.
The fractional r.m.s. variations, calculated from the daily averaged
fluxes, in the six energy bands are 62\% (1.3-3.0 keV),
40\% (3.0-4.8 keV), 17\% (4.8-12 keV), 43\% (20-50 keV), 54\% (50-100 keV)
and 62\% (100-200 keV). As a comparison, the 1.3-200 keV luminosity r.m.s. 
fractional variation is about 15\%.
This behaviour indicates a spectral pivoting at around 10 keV.

To obtain the bolometric luminosity, we need to correct
for the flux emitted beyond both the low-energy (1.3 keV) 
and high-energy (200 keV) limits of our observations. The
high-energy correction is not important due to the rather steep
spectral cutoff around 100 keV in the HS and the rather steep power law
(photon index around --2.5) in the SS. The total luminosity above
200 keV is comparable in both the HS and SS and amounts to
less than 10\% of the bolometric luminosity. 

The low-energy cutoff correction is
necessary because a significant portion of the ultra-soft component 
is not detected with ASM. This component can be fitted with a blackbody (BB).
For the HS, values of $kT_{\rm bb} \sim$ 0.12-0.16 keV were obtained from one
ROSAT (0.1-2.0 keV)
observation on 1991 April 18-20 (Balucinska-Church \etal\
1995) and 11 ASCA (0.5-10 keV) observations between 1993 October and 
1994 December 
(Ebisawa \etal\ 1996). Values of $kT_{\rm bb} \sim$ 0.34 keV were obtained 
in the SS, obtained from one
ASCA observation on 1996 May 30 during the SS (Dotani \etal\ 1996), 
and from 11 RXTE PCA and HEXTE observations
throughout the whole SS (Cui \etal\
1996b). The RXTE data indicate that the BB temperature is slightly higher
in some RXTE observations (Cui \etal\ 1996b). We thus take the 0.34 keV as the
lower limit of the BB temperaturein the SS.
It is not possible to estimate the BB component in the HS from
the ASM data since the power law component dominates the ASM HS
detector counting rates.
Previous ROSAT and ASCA observations of Cyg~X-1 in the HS revealed
a BB component luminosity level of about 5$\times$10$^{36}$ erg/s. We
added this value to the HS 1.3-200 keV luminosity as shown in
the bottom panel of Fig. 1. The soft excess accounts for about 50-70\% of the 1.3-3.0
keV ASM counts in the SS. Taking the BB temperature {\it kT}
$\ga$0.34 keV, we estimate that the BB luminosity below 1.3 keV is  
$\la$1.2-1.7$\times$10$^{37}$ erg/s (the total SS BB luminosity is $\la$
2.2-3.1$\times$10$^{37}$ erg/s). The value of 1.7$\times$10$^{37}$ erg/s
is added to the SS 1.3-200 keV
luminosity as depicted in the bottom panel of Fig. 1. In summary, SS bolometric
luminosity is between the thin and thick curves.

The bolometric luminosity displays an increase and decrease by 
$\la$50-70\%
during the H-to-S and S-to-H state transitions. 
In the middle of the SS, the luminosity is however, only
$\la$10-20\% higher than the HS luminosity, because of
the gradual decrease of the HXR flux after the initial SS
transition and the almost symmetric recovery after reaching
the minimum HXR flux
between TJD 10260 and 10275, while the soft excess in the 1.3-3.0 keV band,
averaged over several days, stayed at about 1.5 Crab during the entire SS.
Therefore the bolometric luminosity variations throughout the state transitions
are less than 50-70\%.

\section{Discussion}

Compared to the high-energy observations of the previous Cyg~X-1 SS
transitions, the data presented here are the first set covering the whole
transition episode continuously over a broad energy band.
The observed 1.3-15 keV ({\it Ariel} V) spectral evolution (spectral pivoting and 
power law index changes) (Chiappetti \etal\ 1981) during the 1975 S-to-H state
transition is quite similar to that presented here. 
The initial HXR flux and spectral steepening observed with HEAO 3 
between 48-183 keV
(Ling \etal\ 1983) 
and the soft X-ray flux level between 1-12 keV observed later
with {\it Hakucho} (Ogawara \etal\ 1981)
are similar to that during the SS presented here. Both previous SS
lasted between 60-100 days, again similar to the 1996 SS duration.
So this 1996 SS
transition is qualitatively similar to the previous ones. It is thus reasonable
to assume that the same physical mechanism is responsible for all of them.

GX~339-4 is the only other BHC observed to have recurrent state transitions.
From a comparison with its 1981 H-to-S state transition (Maejima
\etal\ 1984), we find that the HS spectra of this source and
Cyg~X-1 are quite similar. The ratio
between the SS and HS total luminosity observed from GX~339-4 
(e.g., Ricketts 1983) is also
similar to that of Cyg~X-1 presented here. During a SS
observation of GX~339-4 in 1983 (Makishima \etal\ 1986), the HXR
flux increased significantly while the soft X-ray flux remained nearly unchanged,
also similar to the HXR flux increase in the second half of the
SS of Cyg~X-1. The photon index of the power law changed from
--0.9 to --2.1 following the HXR flux increase of GX~339-4, different from the
near constant value of --2.5 during much of the SS in Cyg~X-1. The power law
tail during a `very high' state of GX~339-4 (Miyamoto \etal\ 1991), when the overall
flux was about a factor of 2-3 higher than during the previous SS
observations, is however, very similar to that of Cyg~X-1 in the SS. 
Qualitatively similar S-to-H state transitions have also been observed
from the low-mass X-ray binary BH system GS~1124--683 (Ebisawa \etal\ 1994)
and the neutron star system 4U~1608--52 (Mitsuda \etal\ 1989).
Therefore similar physical mechanisms might operate in all of them.

The `hard' and `soft' states we refer to in this paper are usually called
the `low' state, and the `high' (and `very high') states. This is 
due to the fact that the
early observations of them were made in the low-energy 
X-ray band ($<$20 keV; see e.g., Tananbaum et al. 1972), so 
the terms `low' and `high' refer to the low and high values of the 
low-energy X-ray fluxes, respectively. It is generally 
believed that the low-energy X-ray flux 
tracks the total mass accretion rate of the system; therefore, these 
`low' and `high' states have been considered to correspond to low and high 
values of \mdot~, respectively. This idea gained strong support
in the unified scheme of source states of both neutron star
and BH X-ray binaries proposed by Van der Klis (1995). 

This picture may be incomplete in view of the lack of a strong 
variation 
of the total luminosity during the state transitions which we observed in 
1996. The blackbody component detected in both the hard 
and soft states is generally thought to originate 
from an optically thick and geometrically
thin accretion disk near the BH (see, e.g., Mitsuda 
et al. 1984). According to current models the hard power law spectral component is likely 
produced in a very hot and optically thin region, through Comptonization of 
low-energy X-ray photons; the detailed nature of the 
optically thin region and the mechanism that makes it very hot 
distinguish these different models. The near constant luminosity during the transitions indicates that they are 
driven by a redistribution of the gravitational energy release 
between the optically thick and
the optically thin regions. Therefore {\it a H-to-S state transition, and its reverse, probably reflects a change in the 
relative importance of the energy release in the optically thin 
and thick regions of the accretion disk near the BH and this may not
require a substantial change in the total accretion rate.} 

Spectra emitted by accretion disks around black holes are well
described by the `multi-temperature disk blackbody model' of Mitsuda
et al. (1984), which has as fit parameters the inner disk radius,
$R_{\rm in}$, and the temperature, $T_{\rm in}$, at that radius. It is
usually assumed that the inner disk radius equals three times the
Schwarzschild radius, and this provides acceptable mass estimates for
the black hole (see Tanaka and Lewin 1995, for a review).

The blackbody fits made to the high-energy tail of this multi-temperature disk
BB model model
provide good fits for a blackbody temperature, $T_{\rm bb}$, which turns out to
be equal to
the temperature in the disk at a radius of about 7 Schwarzschild radii
(Ross et al. 1992); correspondingly, one has $T_{\rm bb} \simeq 0.7
T_{\rm in}$. Independent of the ratio of inner disk radius to the
Schwarzschild radius, one has $T_{\rm bb} \propto T_{\rm in}$. Since
the multi-temperature disk model the bolometric luminosity,
($L_{\rm bol,disk}$) follows the proportionality relation
$L_{\rm bol,disk} \propto R_{\rm in}^2~T_{\rm in}^4$, one has
$(L_{\rm h}/L_{\rm s}) = (R_{\rm h}/R_{\rm s})^2 (T_{\rm h}/T_{\rm
s})^4$, by applying the proportionality relation in both the HS and SS,
where the subscripts `h' and `s' denote the hard and soft states
respectively.

From the observations we infer that $L_{\rm s}/L_{\rm h} \la 6$, and
$T_{\rm s}/T_{\rm h} \ga $ 2.1 to 2.8. Consequently, we obtain
$R_{\rm h}/R_{\rm s} \ga $ 1.8 to 3.2. This result suggests that 
{\it during the H-to-S
transition the inner radius of the geometrically thin and optically
thick disk changed from $\ga 170$ km to $\sim 70$ km} (the latter
value has been obtained by Dotani et al. 1997, from ASCA observations
during the soft state). Applying the relationship between the
mass accretion rate, inner disk radius and the inner disk temperature in
the multi-temperature disk model we have $\dot M_{\rm s}/\dot M_{\rm
h} \la $ 2.0 to 3.4. 

This change in inner disk radius accompanying the spectral state
changes suggests a picture in which during the hard state  advection of
internal disk energy into the black hole, as proposed by Narayan
(1996), dominates within a radial distance of $\ga 170$ km from the
hole. In the soft state the flow in the inner disk may still be
advection dominated, with only a moderate fraction of the mass flow
passing through an optically thick inner disk.

Variations of the inner disk radius have also been discussed by Ebisawa \etal\ 
(1996), in the context of the mass accretion and high-energy radiation model of
Chakrabarti and Titarchuk (1995), and have recently been proposed by Belloni
et al. (1996) as an explanation for the rapid variability of the
black-hole candidate GRS~1915+105 (albeit at much higher mass
accretion rates).

A potential problem for any model that purports to explain the
spectral transitions in terms of an instability in the inner disk
region is that the transitions seem to be accompanied by the gradual
changes in the slope and the flux of the HXR power law component, 
which occur over
a very long time scale ($\sim$ weeks). In fact, the 1996 S-to-H
transition was predicted from the HXR flux increase (Zhang et al.
1996c). It is possible that Cyg~X-1 underwent a very slow change in
the mass accretion rate, which initially only affected the properties
of the hard X-ray emission region. 
The sudden soft (1.3-3.0 keV) X-ray flux increase and decrease over a much 
shorter time scale, during the state transitions, may then
represent the crossing of a threshold, at which the radiative
efficiency exceeds a value required for the formation of an optically
thick disk down to the innermost stable orbit.

In view of the observational limitations the above remarks are
necessarily somewhat speculative. It would appear that a better
understanding of the cause of the spectral state changes in Cyg X-1
requires that further monitoring of the long-term behaviour of Cyg X-1
include low-energy ($\leq 1$ keV) coverage of the source as well.

\acknowledgements{We thank the ASM/RXTE and BATSE/CGRO teams for
providing excellent technical supports. SNZ thanks W. Chen, K. Ebisawa, 
L. Titarchuk, 
S. Chakrabarti, R. Narayan and C. Robinson for many stimulating
discussions. JvP acknowledges support from NASA through grant NAG5-3003.
We also appreciate very much the constructive comments and
suggestions from the anonymous referee.}

\newpage
\figcaption[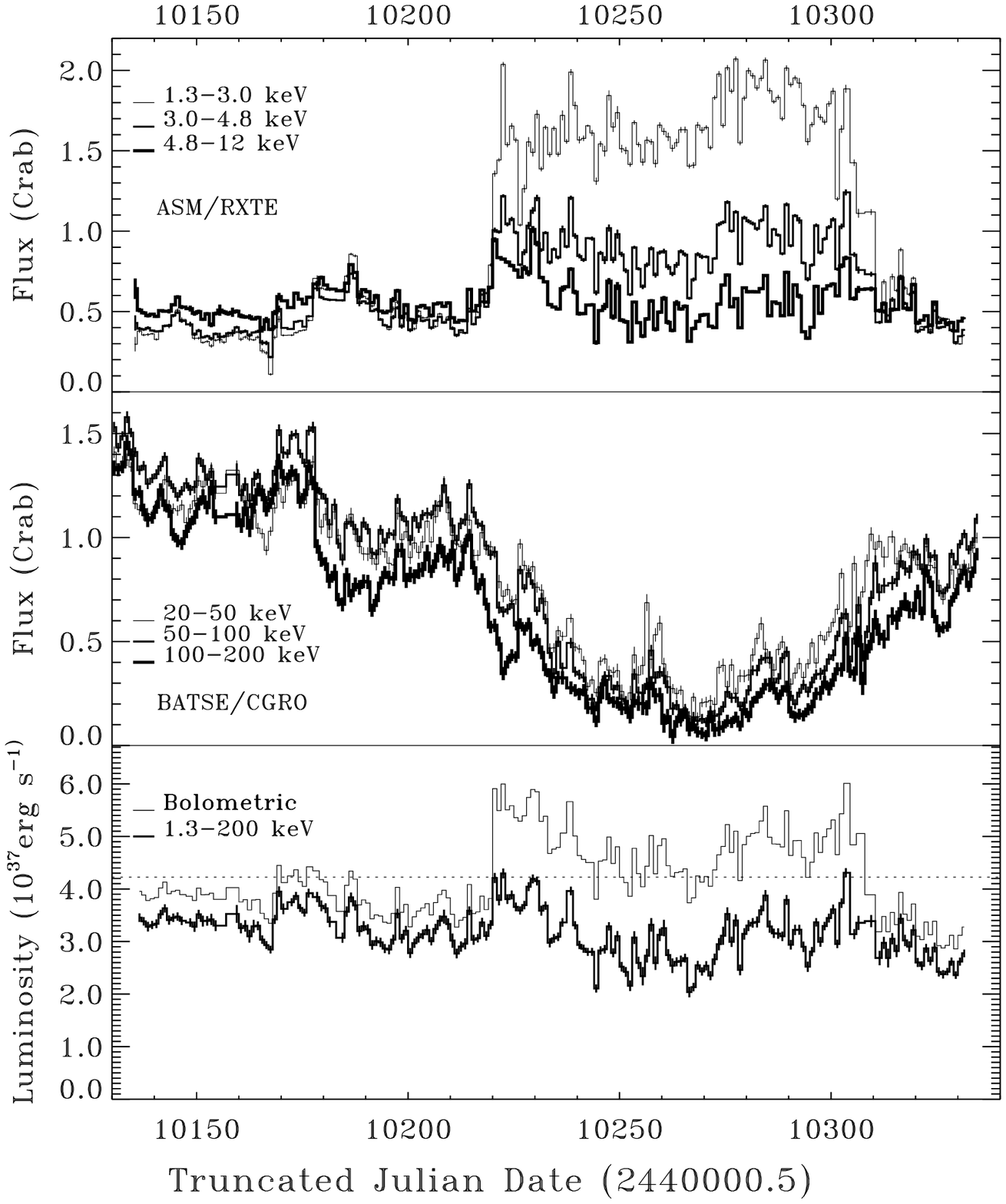]
{Cyg~X-1 light curves and the 1.3-200 keV luminosity variations
during its 1996 SS transition.
\label{figure1}
}

\newpage
\setcounter{figure}{0}

\begin{figure}
\centerline{
\psfig{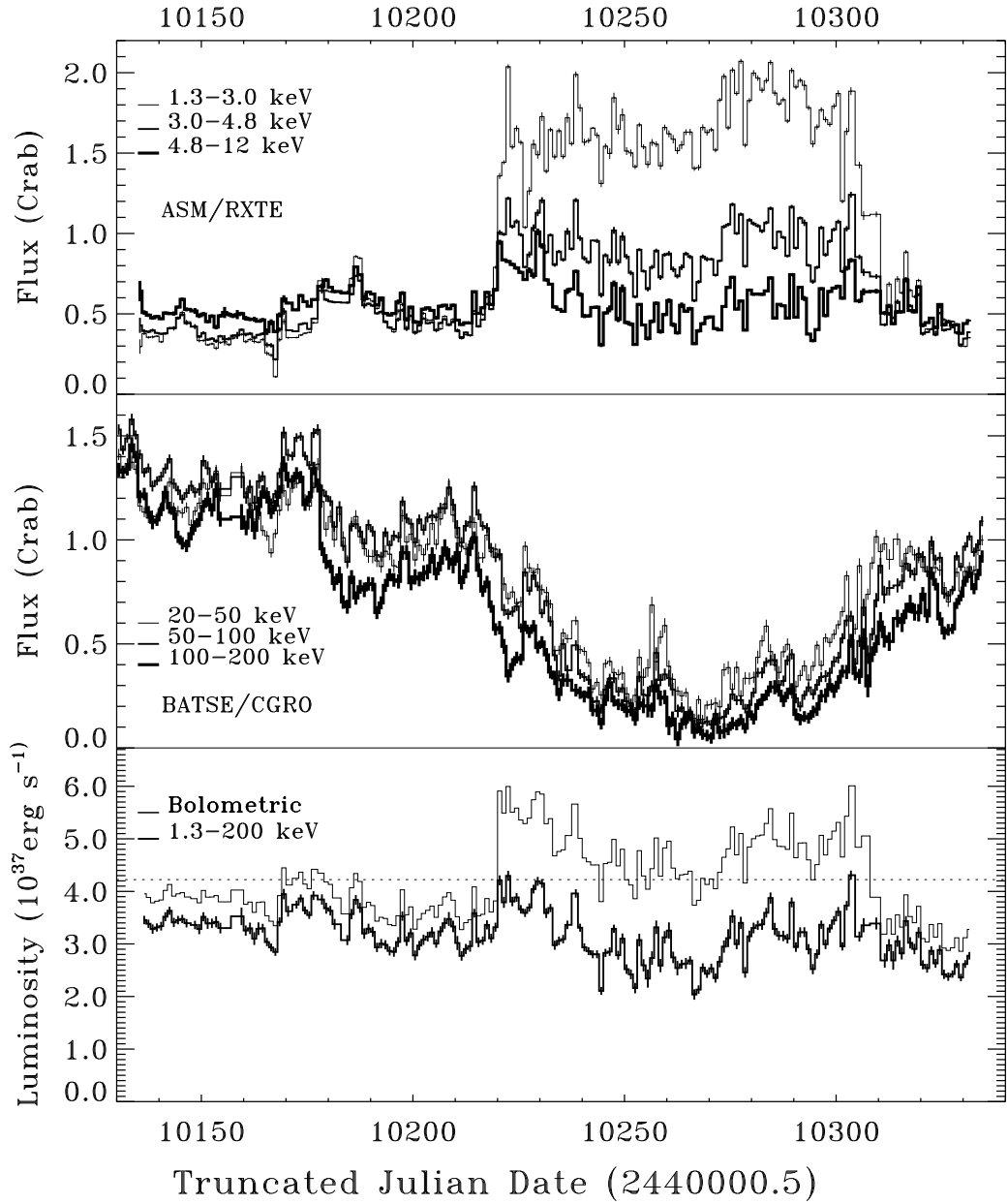}
}
\caption{Cyg~X-1 light curves and luminosity variations
during its 1996 SS transition.}
\end{figure}

\end{document}